%% file: ijcai25.tex
\documentclass{article}
\usepackage{ijcai25}

\usepackage{times}
\usepackage{soul}
\usepackage{url}
\usepackage[hidelinks]{hyperref}
\usepackage[utf8]{inputenc}
\usepackage[small]{caption}
\usepackage{graphicx}
\usepackage{amsmath}
\usepackage{amsthm}
\usepackage{booktabs}
\usepackage{algorithm}
\usepackage{algorithmic}
\usepackage[switch]{lineno}
\usepackage{enumitem}
\usepackage{xcolor}
\newcommand{\model}{PS\textsuperscript{2}\xspace}
\input{math_commands}

\usepackage{xspace}
\usepackage{tcolorbox}
\usepackage{multirow}
\usepackage{array}
\usepackage{subfigure}
\usepackage{pdfpages}
\usepackage{verbatim}
\usepackage[export]{adjustbox} 
\usepackage{marvosym}
\usepackage{bm}
\usepackage{footmisc}
\usepackage{subcaption}
\usepackage{color}
\usepackage{colortbl}
\usepackage{tabularx}
\usepackage{rotating}
\usepackage{graphicx}
\usepackage{subcaption}  
\usepackage{lipsum}      
\usepackage[table]{xcolor}
\usepackage{tcolorbox}
\tcbuselibrary{breakable}
\usepackage{multicol}
\usepackage{xcolor}
\usepackage{listings}
\usepackage{xcolor}

\newtcolorbox{promptbox}{
  breakable,
  colback=gray!5,
  colframe=gray!60,
  title=Prompt,
  fonttitle=\bfseries,
  left=6pt,
  right=6pt,
  top=6pt,
  bottom=6pt
}

\title{PS\textsuperscript{2}: Parameterized Control for Fine-Grained Student Proficiency Simulation}

\author{
Ruochen Liu\footnotemark[1]$^1$\and
Zhiyuan Wen\footnote{Equal Contribution}$^2$\and
Hao Yan$^{1}$\and
Jun Yin$^1$\and
Senzhang Wang$^1$ $^{\textrm{\Letter}}$\and
Jiannong Cao$^2$
\\
\affiliations
$^1$Central South University\\
$^2$The Hong Kong Polytechnic University\\
\emails
\{ruochen, csuyh1999, szwang\}@csu.edu.cn,
\{zyuanwen, jiannong.cao\}@polyu.edu.hk,
\{Junmay.yin\}@connect.polyu.hk,
}

\begin{document}
\maketitle

\begin{abstract}
\input{Content/absv2}
\end{abstract}

\section{Introduction}

\input{Content/introv2}

\section{Related Work}
\input{Content/related-work}
\section{Preliminaries}

\input{Content/pre2}

\section{Methodology}
\input{Content/Method}

\section{Experiments}
\input{Content/experiments}

\section{Conclusion}
In this paper, we address the limitations of prompt-based student proficiency simulation, which often yields coarse control. To better model cognitive differences, we propose \model, a model-level framework that parameterizes proficiency via logits-level interpolation between upper- and lower-bound student LLMs. The lower-bound model is trained with cognitively informed errors to generate plausible low-proficiency behaviors, and \model calibrates the hybrid ratio to align simulated students with target proficiency levels. Experiments on two public datasets show that \model outperforms existing baselines in both behavior similarity and difficulty prediction.
\clearpage

\bibliographystyle{named}
\bibliography{ijcai25}
\clearpage
\newpage
\appendix
\input{Content/appendix}

\end{document}

%% file: math_commands.tex
\usepackage{amsmath,amsfonts,bm}

\def\eqref#1{equation~\ref{#1}}

\def\1{\bm{1}}

\def\vz{{\bm{z}}}

\DeclareMathAlphabet{\mathsfit}{\encodingdefault}{\sfdefault}{m}{sl}
\SetMathAlphabet{\mathsfit}{bold}{\encodingdefault}{\sfdefault}{bx}{n}



%% file: Content/absv2.tex
Understanding how students with different proficiency levels respond to educational materials is a critical issue within the field of AI for Education. However, acquiring sufficient real student response data for a robust evaluation is often hindered by cost, ethics, and security constraints. Consequently, LLM-based student proficiency simulation, especially prompt-based methods, has emerged as a practical alternative under data-scarce conditions. Despite their promise, current methods still exhibit limited controllability with coarse-grained proficiency representations, high sensitivity to prompt design, and the lack of calibration with academic performance. Therefore, we propose \textbf{P}arameterized \textbf{S}tudent \textbf{P}roficiency \textbf{S}imulation (\textbf{\model}), an unsupervised and parameterized model-level framework that simulates students with different proficiencies by interpolating between a strong upper-bound LLM and a weaker, cognitive error-informed lower-bound student LLM via a hybrid ratio.
Specifically, the lower-bound model is constructed by fine-tuning the weaker LM to exhibit cognitive errors when responding to educational materials. To ensure alignment with target proficiency levels, \model further calibrates the interpolation ratio with academic performance. Experiments on two public datasets demonstrate that \model achieves finer-grained and consistent proficiency simulation compared to existing baselines, leading to superior performance in student behavior similarity and item difficulty prediction.

%% file: Content/introv2.tex
Understanding how students at varying proficiency levels respond to educational materials is a critical challenge in educational research and intelligent tutoring systems \cite{corbett1994knowledge}. Such responses provide valuable insights for assessing question difficulty and guiding instructional interventions \cite{piech2015deep,abdelrahman2023knowledge}. In practice, however, acquiring large-scale student responses is often time-consuming and costly, especially in dynamic learning environments with rapidly evolving content \cite{baker2016educational,choi2020ednet}. Consequently, student proficiency simulation has emerged as a viable alternative, enabling educational systems to evaluate new questions and instructional strategies.
\begin{figure}[t]
\centering

\includegraphics[width=1.02\linewidth]{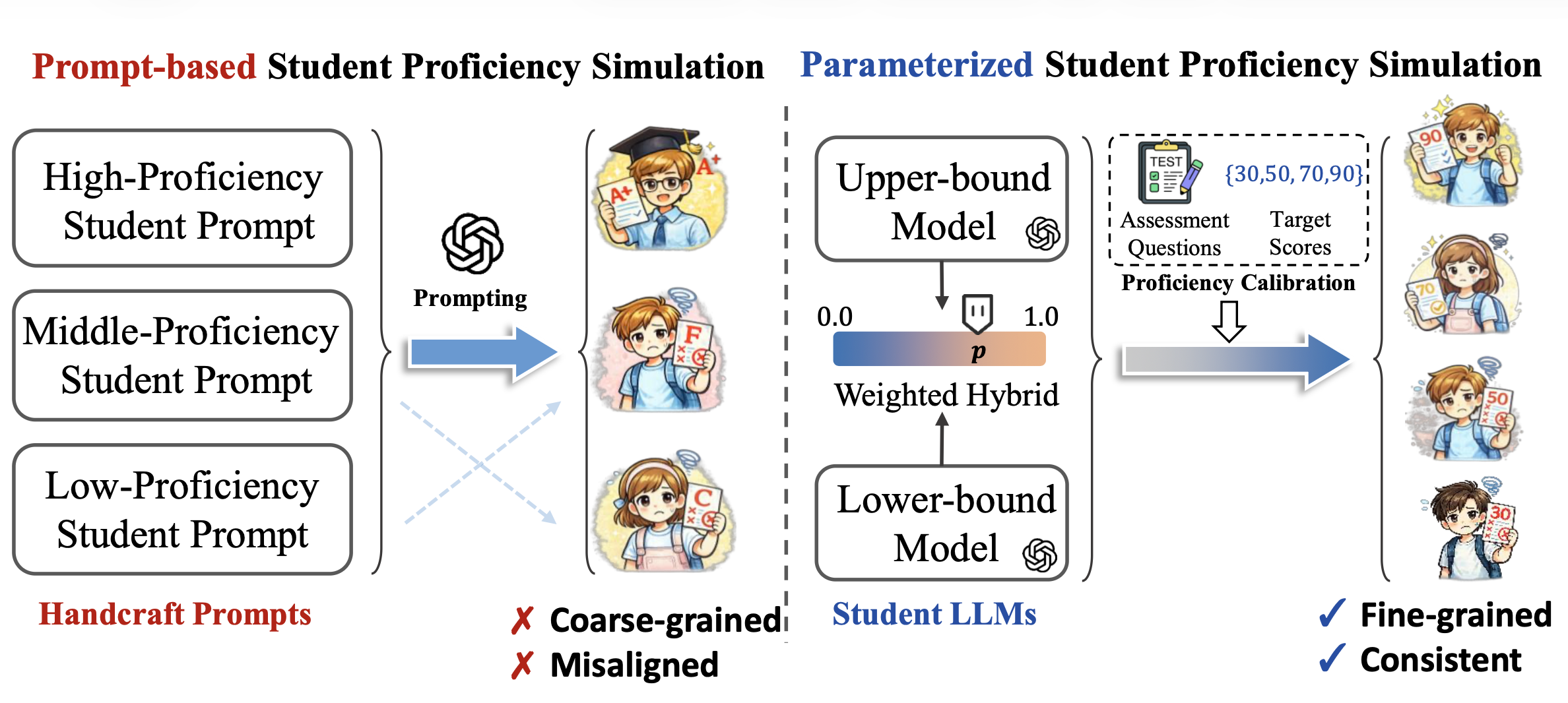}

\caption{{Prompt-based Student Proficiency Simulation v.s. Parameterized Student Proficiency Simulation.
}}
\label{fig:intro}
\end{figure}

Traditional approaches to student proficiency simulation train models on real student responses annotated with performance indicators, such as assessment performance or Item Response Theory (IRT)-based difficulty estimates~\cite{rasch1993probabilistic,edelen2007applying}, and supervise models to generate proficiency-aligned responses. Prior work has explored both multiple-choice and open-ended settings by aligning model outputs with real student responses~\cite{feng2025reasoning}. However, privacy concerns and device limitations often hinder the acquisition of large-scale data, thereby restricting the applicability of these methods~\cite{zeide2015student,marin2025ethical}. Given that LLMs are trained on vast corpora of human-generated data~\cite{zhao2023survey,khan2024ethical}, recent research has investigated prompt-based student proficiency simulation as a promising unsupervised alternative. For example, \cite{benedetto-etal-2024-using} defines distinct proficiency levels and instructs LLMs to answer questions accordingly, while others~\cite{lu2024generative,minaee2024large} enhance simulation fidelity by incorporating proficiency-specific knowledge directly into the prompts.

While prompt-based simulation offers an unsupervised approach to modeling proficiency, it provides limited control over the reasoning alignment with target levels.
As shown in the left part of Figure~\ref{fig:intro}, existing prompt-based approaches have the following limitations:
(i) \textbf{Coarse-grained student proficiency simulation.} 
Text prompts have difficulty precisely capturing the distinctions between fine-grained proficiency levels. Once proficiency levels become finer, response distributions may overlap. Lower-proficiency models may sometimes outperform higher-proficiency ones~\cite{williamson2023cognitive}. 
(ii) \textbf{Sensitive in designing effective prompts.} Small changes in prompts can lead to large differences in model responses, making it hard to ensure a consistent and reliable simulation across proficiency levels. (iii) \textbf{Lack of score-based calibration.} 
Student proficiency is usually measured by assessment scores, and mapping scores to concrete proficiency descriptions requires rigorous expert knowledge. Without such mappings, prompt-based methods cannot reliably map simulated students to target proficiency levels.

Instead of relying on prompts, we argue that effective proficiency simulation requires modeling differences in cognitive capabilities. We therefore simulate students at different proficiency levels by directly controlling model performance through a parameterized mechanism. A straightforward strategy is to treat a high-performing LLM as an upper-bound student and perturb its outputs with noise, using noise strength as a control parameter to match a target lower proficiency level. However, this approach is non-trivial to realize due to the two major challenges:
(i) \textbf{Undirected performance degradation}. Simple perturbation risks impairing the model's basic language modeling capability, causing simulated low-proficiency students to produce random outputs rather than reasonable errors. 
(ii) \textbf{Uncalibrated perturbation magnitude} There is no empirical standard for setting perturbation strength, making it difficult to pinpoint the appropriate noise level for a specific target proficiency. Consequently, reliably generating students at specific proficiency levels necessitates extensive trial and error.

To address the above challenges, we propose \textbf{P}arameterized \textbf{S}tudent \textbf{P}roficiency \textbf{S}imulation (\textbf{\model}), a model-level approach for simulating students with different proficiency levels in fine-grained without real student response data. Instead of using unstructured noise, \model parameterizes student proficiency via a hybrid ratio, blending a high-performing upper-bound student LLM with a lower-bound student LLM that produces cognitive errors at the logits level for decoding. By interpolating between these two models, \model generates students across a continuous proficiency spectrum while maintaining basic language ability. To obtain such a lower-bound LLM, we introduce a cognitive error generation strategy that explicitly models common conceptual and procedural errors~\cite{mccormick1997conceptual} in student behaviors and uses them to generate training data that guide the model toward cognitive error patterns. Finally, \model calibrates hybrid ratios to align simulated students with target proficiency levels in both score-aware and score-agnostic settings. Our main contributions are summarized as follows:
\begin{itemize}[leftmargin=*]
    \item We propose \model, a model-level framework that parameterizes proficiency by hybridizing upper- and lower-bound student LLM, enabling fine-grained, continuous simulation without real student response data.
    \item We design a cognitively informed error modeling strategy to generate coherent, structured errors instead of random noise, ensuring plausible low-proficiency simulations.
    \item We validate \model on two public datasets, showing it outperforms existing baselines in student behavior similarity and item difficulty prediction, demonstrating finer-grained and more consistent proficiency simulation.
 The source code is available at \textcolor{blue}{{\url{https://anonymous.4open.science/r/StudentSimulation-EDBF}}}.
\end{itemize}

%% file: Content/related-work.tex


Modeling and simulating student proficiency is important in educational research~\cite{marquez2025simulating,vanlehn2006behavior}, as it enables low-cost feedback on instructional content and assessment design.

Most existing student proficiency simulation methods rely on supervised learning, training models on historical student responses to predict answers to new questions. \cite{zelikman-etal-2023-generating} fine-tunes LLMs to reproduce students’ answering behaviors and response times for difficulty prediction. Beyond correctness prediction, \cite{he2024psychometric} infers incorrect answers and aligns the generated response distribution with real data via KL divergence, while \cite{scarlatos2025smart} models open-ended responses under an IRT framework and uses DPO \cite{DPO} to match student preference pairs. Knowledge editing techniques can also be used to edit pretrained models’ behavior on historical responses, enabling student proficiency simulation without full retraining~\cite{sinitsin2020editable,mitchell2021fast}.

Nevertheless, student response data are often difficult to obtain due to privacy concerns or device limitations, limiting the applicability of these methods. Leveraging the fact that LLMs are pre-trained on extensive human-generated text, recent studies have turned to prompt-based student proficiency simulation as an unsupervised alternative. For instance, \cite{benedetto-etal-2024-using} instruct models to answer questions under a set of predefined proficiency levels, while \cite{lu2024generative} enhance prompts with proficiency-related knowledge to generate more realistic student responses.

%% file: Content/pre2.tex


In this section, we first formalize the student proficiency simulation problem. Since prompt-based student simulation is the primary baseline in our study, we then provide an overview of this paradigm before presenting our method.
\subsection{Problem Definition}
Student proficiency simulation aims to build models that, given proficiency descriptions, generate responses similar to real students at the corresponding levels for assessment items.(i.e., test questions or exercises).

Formally, let $\{L_1, \dots, L_N\}$ denote $N$ descriptions of proficiency levels; these descriptions may be expressed in natural language or derived from students' assessment scores. The objective is to build student models $\{S_1, \dots, S_N\}$ such that each model $S_n$ exhibits behavior closely matching that of real students at proficiency level $n$. The student models can then be used to simulate student responses for downstream educational tasks (e.g., item difficulty prediction).

Let $\mathcal{R}_n$ denote the set of reference responses collected from real students at proficiency level $n$. The objective is formalized through a behavior consistency measure
\[
C(S_n, \mathcal{R}_n),
\]
which quantifies how closely the responses generated by the student model $S_n$ align with the reference responses $\mathcal{R}_n$. The measure can be instantiated, for example, using predictive response distribution similarity.

\subsection{Prompt-based Student Proficiency Simulation}
Prompt-based student simulation leverages LLMs to generate responses corresponding to different proficiency levels, guiding the model with carefully designed prompts. Given an item $i$ and a target proficiency level $n$, a student response is generated as
\begin{equation}
\mathbf{x}_{i}^{(n)} = \operatorname{Prompt}(i, n), \quad
S_n(i) = \operatorname{LLM}(\mathbf{x}_{i}^{(n)}, \bm{\theta}),
\end{equation}
where $\mathbf{x}_{i}^{(n)}$ encodes the question along with a description of the target proficiency level $n$, $S_n(i)$ denotes the response generated by the student model $S_n$, and $\bm{\theta}$ represents the parameters of the LLM.

In practice, however, prompt-based simulation depends on manually crafted prompts, which provide limited and sometimes inconsistent control over student proficiency.


%% file: Content/Method.tex
\begin{figure*}[t]
\centering
    \includegraphics[width=\linewidth]{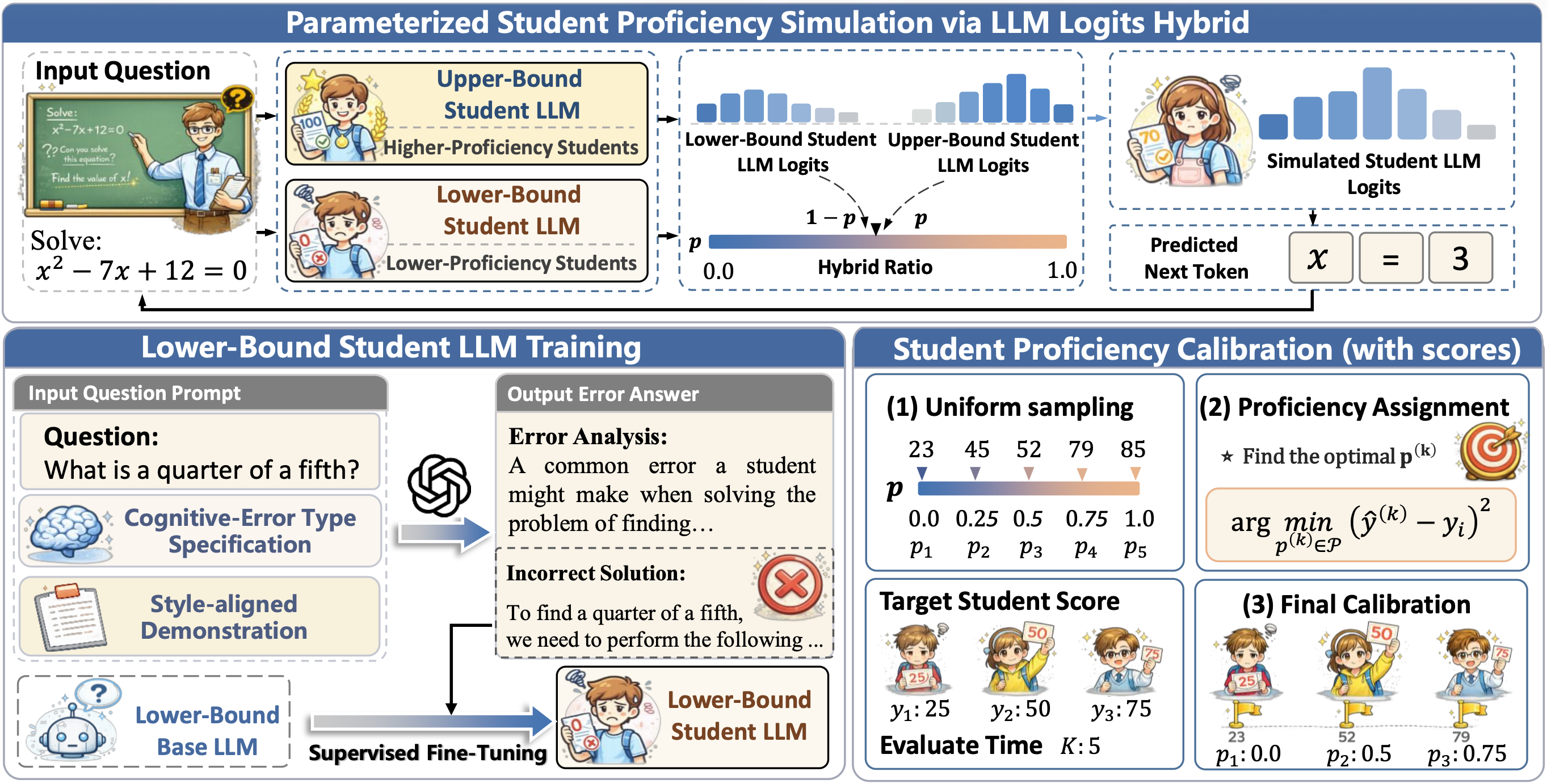}
    \caption{The top panel shows \model simulating students by hybridizing upper- and lower-bound LLMs at the logits level, while the bottom panels depict lower-bound training with cognitive errors and calibration of hybrid ratios using assessment scores.}
\label{fig:framework}
\end{figure*}

In this section, we describe the design of \model for simulating students across varying proficiency levels. We first introduce the upper- and lower-bound modeling framework, then describe how the lower-bound student LLM is trained using a cognitively informed error modeling strategy, and finally present how student proficiency is calibrated when target assessment scores are available. The overall framework is illustrated in Figure~\ref{fig:framework}.

\subsection{\model Framework}

We first formally introduce the core modeling framework of \model, which defines how student proficiency is controlled.

Our core idea is to simulate students across proficiency levels by interpolating between two models: an upper-bound student LLM representing high-performing students and a lower-bound student LLM representing low-performing students. This interpolation enables generating students at intermediate proficiency levels. The upper-bound student LLM is an existing pre-trained LLM, while the lower-bound student LLM is trained as described in Section~\ref{Lower-Bound Model Training}.

We interpolate the models at the logit level. Formally, let $\vz_{i,t}^{u}$ and $\vz_{i,t}^{l}$ denote the logits of the upper-bound
and lower-bound LLMs at step $t$ for item $i$, given previously generated tokens
$\mathbf{x}_{<t}$. For a student at hybrid ratio $p \in [0,1]$, the logits are
linearly interpolated as
\begin{equation}
    \vz_{i,t}(p) = (1-p)\, \vz_{i,t}^{l} + p \, \vz_{i,t}^{u}.
\end{equation}

The corresponding conditional token distribution is then given by
\begin{equation}
    \pi(x_t \mid \mathbf{x}_{<t}, i, p) = \text{softmax}\big(\vz_{i,t}(p)\big),
\end{equation}
and the full sequence distribution is
\begin{equation}
    \pi(\mathbf{x}_{\le T} \mid i, p) = \prod_{t=1}^{T} \pi(x_t \mid \mathbf{x}_{<t}, i, p).
\end{equation}
By assigning different hybrid ratios $p$ to different target proficiency levels, we can thus obtain simulated students with varying proficiency levels.

Notably, for $p \in [0,1]$, the KL~\cite{kullback1951information} divergence between the output distribution of the simulated student model and that of the upper-bound student LLM,
\[
    \mathrm{KL}\!\left(\pi(\cdot \mid i, p)\,\|\,\pi_{\mathrm{u}}(\cdot \mid i)\right),
\]
is monotonically increasing with respect to $p$ (see Appendix \ref{Appendix:Proof} for a proof). This monotonicity ensures that the simulated student’s behavior becomes progressively closer to that of the upper-bound student LLM as $p$ increases \cite{gou2021knowledge,park2019relational,cho2019efficacy}, providing a theoretical basis for assigning different $p$ values to target proficiency levels (see Section \ref{Calibration}).

\subsection{Lower-Bound Student LLM Training}
\label{Lower-Bound Model Training}
To simulate students at different proficiency levels, we design a lower-bound student LLM that produces coherent but low-proficiency responses. This model serves as a stable reference baseline for low-level student proficiency. We fine-tune it using a cognitive error generation strategy without relying on real student responses, enabling it to generate characteristic student errors while maintaining basic linguistic fluency and reasoning coherence.

\subsubsection{Cognitive Error Generation Strategy.} 
We generate cognitive erroneous solutions by modeling two common types of cognitive mistakes \cite{mccormick1997conceptual,rittle2014developing}. Our prompting strategy consists of two parts: a cognitive error specification that defines the error types, and a style-aligned demonstration that preserves the format of the base model.

\paragraph{\textbf{\textit{Cognitive Error Specification.}}}
The cognitive error specification defines the types of student mistakes that the model is expected to generate. We consider two main categories of errors: conceptual errors and procedural errors.

Conceptual errors arise from misunderstandings of underlying principles or problem semantics. 
\begin{tcolorbox}[title = {Conceptual-Error Specification}]
\textbf{Conceptual Error:}
A student exhibits a systematic misunderstanding of relevant knowledge, concepts, principles, or relationships when solving a problem. This leads to a fundamental error in the reasoning chain. Even if the computational steps are correct, the final result is still incorrect.
\end{tcolorbox}
Procedural errors arise from incorrect execution of solution steps rather than conceptual misunderstandings.

\begin{tcolorbox}[title = {Procedural-Error Specification}]
\textbf{Procedural Error:}
An error that occurs when a student applies knowledge, skills, or strategies incorrectly during problem-solving or task completion. Such errors may arise at different levels—routine skills, strategic operations, or higher-order control processes. Even if the student understands the underlying concepts correctly, procedural errors can lead to incorrect outcomes.
\end{tcolorbox}

\paragraph{\textbf{\textit{Style-aligned demonstration.}}} 
To ensure that the lower-bound model generates errors without deviating excessively from the original base model in format or style, we include a style-aligned demonstration in the prompt to guide consistent formatting and presentation.

Specifically, given an item $i$ and the base lower-bound student LLM $M_0$, the demonstration answer produced by the base model is denoted as:
\begin{equation}
a_i^{\mathrm{base}} = M_0(q),
\end{equation}
where $a_i^{\mathrm{base}}$ represents the demonstration generated by $M_0$.

To generate error-specific answers, we employ an error generator $\mathcal{G}$ (e.g., ChatGPT \cite{achiam2023gpt}). For each error type $e \in \{\text{conceptual}, \text{procedural}\}$, an erroneous answer $a_i^{(\mathrm{err}, e)}$ is generated as:
\begin{equation}
a_i^{(\mathrm{err}, e)} = \mathcal{G}\big(i, e, a_i^{\mathrm{base}}\big),
\end{equation}

The detailed prompt used for error generation is provided in Appendix~\ref{Appendix:prompt}.

\subsubsection{Lower-Bound Student LLM Fine-Tuning}
Using the above procedure, we construct a fine-tuning dataset consisting of item–answer pairs that reflect structured low-proficiency proficiency:
\begin{equation}
\mathcal{D}_{\mathrm{LB}} = \{ (i_n, a_{i_n}^{\mathrm{(err, e)}}) \}_{n=1}^{N}.
\end{equation}
We then fine-tune the base lower-bound student LLM $M_0$ on $\mathcal{D}_{\mathrm{LB}}$ using supervised learning, encouraging the model to internalize cognitive error patterns while preserving basic linguistic coherence. Concretely, the fine-tuning objective is defined as:
\begin{equation}
\mathcal{L}_{\mathrm{SFT}}(\theta) 
= - \mathbb{E}_{(i,a)\sim \mathcal{D}_{\mathrm{LB}}}
\sum_{t=1}^{|a|}
\log p_\theta \left(a_t \mid i, a_{<t} \right),
\end{equation}
where $\theta$ denotes the trainable parameters of the base lower-bound LLM.
To improve parameter efficiency and training stability, we adopt LoRA \cite{hu2021lora} for fine-tuning.

\subsection{Student Proficiency Calibration}
\label{Calibration}

In this section, we discuss how to assign a hybrid ratio 
$p$ to each proficiency level. We consider two scenarios: no target scores and with target scores. Since the score of the simulated student increases monotonically with 
$p$, we employ uniform sampling to cover the entire proficiency space and design calibration strategies for each scenario based on this sampling.

\paragraph{(1) No target scores.}
When no target score is available, we only need to simulate students at different proficiency levels.
Let $N$ be the number of desired levels. We sample $N$ proficiency values uniformly in $[0,1]$:
\[
p_i = \frac{i}{N}, \quad i = 1, \ldots, N.
\]

\paragraph{(2) With target scores.}
When target proficiency scores $\{y_1, \ldots, y_N\}$ on item set $\{i_1,\ldots,i_M\}$ are available, we sample $K \ge N$ proficiency values uniformly to improve calibration accuracy:
\[
p^{(k)} = \frac{k}{K}, \quad k = 1, \ldots, K.
\]
We evaluate the model on the same item set at each $p^{(k)}$ to obtain predicted accuracies $\hat{y}^{(k)}$, and then assign each target score $y_n$ to the closest sampled score:
\[
p_n = \arg\min_{k} \left( \hat{y}^{(k)} - y_n \right)^2, \quad n = 1, \ldots, N.
\]
The selected $\{p_1, \ldots, p_N\}$ are used for simulation.

%% file: Content/experiments.tex
In this section, we conduct extensive experiments aiming to answer the following questions.
\textbf{RQ1:} How effectively can \model support a downstream evaluation task by modeling student response behavior? 
\textbf{RQ2:} To what extent can \model generate student responses whose distribution is consistent with real student data? 
\textbf{RQ3:} Can \model reliably control and reflect different student proficiency levels under varying granularity settings, and does it preserve the correct ordering between student proficiency?
\textbf{RQ4:} How key components of \model affect its performance?

\subsection{Experimental Setup}
\subsubsection{Datasets}
To assess our method, we conduct experiments on two real-world student response datasets: the multiple-choice EEDI dataset and the open-ended SciEntsBank dataset.

{\textbf{EEDI} \cite{pmlr-v133-wang21a}.} EEDI is derived from the NeurIPS 2020 Education Challenge dataset from the Eedi platform, containing student responses to math multiple-choice questions. We extracted text from the original question images and removed items requiring visual reasoning (e.g., graphs or diagrams). The resulting dataset includes 422 questions and 443,433 responses from 5,533 students, split 80:20 into training and test sets. IRT was applied separately to each split to estimate question difficulty, which was used as the ground-truth ranking for evaluation.

\textbf{SciEntsBank} \cite{dzikovska-etal-2013-semeval}. The SciEntsBank dataset contains ~11,000 short-answer responses to 197 science questions across 15 domains. We use the 2-way labels and evaluate on the Unseen Domain (UD) and Unseen Questions (UQ) splits. As user identifiers are unavailable, questions are ranked by overall accuracy as ground truth. Following \cite{duan2025}, we train a scoring model to assess response quality; training details are provided in Appendix~\ref{Appendix:score_model}.

Since the EEDI dataset only provides students’ selected answer choices, we conduct Student Simulation experiments exclusively on SciEntsBank. Question difficulty prediction is evaluated on both datasets.

\subsubsection{Metrics}

\paragraph{Question Difficulty Prediction.}
Following prior work in question difficulty prediction \cite{AAAI2017Huang,he2024psychometric}, we use the widely adopted Pearson Correlation Coefficient (\textbf{PCC}) to measure the linear correlation between predicted and ground-truth item difficulties, Spearman Rank Correlation Coefficient (\textbf{SCC}) to measure the correlation between ranks of predicted and ground-truth item difficulties.

\paragraph{Student Proficiency Simulation.}
Following prior work in population simulation \cite{bui2025mixture,yu2023large}, we compare the generated and ground-truth response distributions in a latent embedding space. We first extract embeddings using the sentence encoder multi-qa-mpnet-base-dot-v1 (Reimers and Gurevych, 2019), then compute Fréchet Inception Distance (\textbf{FID}) \cite{heusel2017gans} and \textbf{MAUVE} \cite{pillutla2021mauve} to assess semantic distribution alignment, and \textbf{Div. KL} to measure the similarity of diversity distributions via pairwise response cosine similarity. Detailed formulas for the metrics are provided in the Appendix \ref{Appendix:Formulation}.

\subsubsection{Baselines}
We evaluate student simulation in an unsupervised setting by comparing the following baselines. \textbf{Vanilla LLM} uses a fixed prompt and samples multiple responses without any prompt modification. \textbf{Conditioned Prompt LLM}~\cite{benedetto-etal-2024-using} partitions students into several proficiency levels via prompt instructions and prompts the model to answer each question as a student with a specified proficiency. \textbf{Knowledge Prompt LLM}~\cite{lu2024generative,minaee2024large} extracts knowledge points from the training set and controls the simulated student’s proficiency by varying the proportion of knowledge provided in the prompt during inference. The detailed implementation is provided in Appendix~\ref{Appendix:Knowledge_Prompt}.

For question difficulty prediction, all baselines use the average score to rank question difficulty. Their output sequences are used to evaluate the quality of student simulation.

\subsubsection{Implementation Settings}
We adopt Llama3.2-3B-Instruct and Llama3.2-8B-Instruct as the upper-bound LLMs for both our simulated student and baseline experiments, while the lower-bound student LLM is Llama3.2-1B-Instruct. GPT-4o is used as the error generator.

On EEDI, we evaluate both {\textbf{with target scores}} and {\textbf{no target scores}}, where target scores are computed as the average proficiency-level scores in the training set. SciEntsBank lacks student IDs and is therefore evaluated only under the no target scores setting. For all baselines, we report the best performance across 3, 5, and 7 proficiency levels. Baseline prompts and hyperparameters are provided in Appendix~\ref{Appendix:prompt} and \ref{Appendix:Implementation}.
\subsection{Question Difficulties Prediction (RQ1)}

\begin{table*}[htbp]
  \centering
  \caption{Performance on question difficulty prediction. Best performance is in \textbf{bold} and second best is \underline{underlined.}}
    \begin{tabular}{lcc|cc|cc}
    \toprule
    \multicolumn{1}{c}{\multirow{2}[4]{*}{Model}} & \multicolumn{2}{c}{EEDI} & \multicolumn{2}{c}{SciEntsBank (UD)} & \multicolumn{2}{c}{SciEntsBank (UQ)} \\
\cmidrule{2-7}
          & PCC$\uparrow$ & SCC$\uparrow$ & PCC$\uparrow$ & SCC$\uparrow$ & PCC$\uparrow$ & SCC$\uparrow$ \\
    \midrule
    \rowcolor[rgb]{ .906,  .902,  .902} \multicolumn{7}{c}{Llama3-3B-Instruct} \\
    Vanilla LLM
      & 0.3094 & 0.2956
      & 0.1732 & 0.2418
      & 0.0189 & 0.0821 \\
    Knowledge Prompt LLM
      & 0.1142 & 0.1598
      & 0.1629 & 0.0632
      & 0.0766 & 0.1412 \\
    Conditioned Prompt LLM
      & 0.3781 & 0.3489
      & \textbf{0.2661} & 0.2188
      & 0.1782 & \underline{0.1862} \\
    \model(w/o target score)
      & {0.4310} & {0.4385}
      & \underline{0.2168} & 0.2574
      & \textbf{0.2770} & \textbf{0.3128} \\
    \model(w target score)
      & \underline{0.4412} & \underline{0.4634}
      & - & -
      & - & - \\
    \rowcolor[rgb]{ .906,  .902,  .902} \multicolumn{7}{c}{Llama3-8B-Instruct} \\
    Vanilla LLM
      & 0.3036 & 0.2807
      & 0.1185 & -0.0627
      & 0.0189 & 0.0821 \\
    Knowledge Prompt LLM
      & 0.1735 & 0.1635
      & 0.1849 & 0.2106
      & 0.1186 & 0.1381 \\
    Conditioned Prompt LLM
      & 0.3575 & 0.3124
      & 0.1754 & \textbf{0.2807}
      & 0.0434 & 0.0110 \\
    \model(w/o target score)
      & 0.4038 & 0.4285
      & 0.2125 & \underline{0.2770}
      & \underline{0.1882} & 0.1644 \\
    \model(w target score)
      & \textbf{0.4495} & \textbf{0.4717}
      & - & -
      & - & - \\
    \bottomrule
    \end{tabular}
  \label{tab:difficulty_prediction}
\end{table*}

To evaluate the practical effectiveness of our approach, we adopt question difficulty prediction as the primary downstream task. The comparative results against other unsupervised baselines are reported in Table~\ref{tab:difficulty_prediction}.

\textbf{\model and its variant without question input consistently achieve the best or second-best performance across all settings.} On Llama3-3B-Instruct, \model attains a PCC of 0.4412 and an SCC of 0.4634 on EEDI, outperforming Vanilla LLM and prompt-based baselines. On the UQ subset, the variant without calibration reaches 0.2770 PCC and 0.3128 SCC, about 0.1 higher than the Conditioned Prompt LLM, indicating stronger uncertainty-aware difficulty modeling. With Llama3-8B-Instruct, \model further improves to 0.4495 PCC and 0.4717 SCC, demonstrating good scalability and robustness.

\textbf{Prompt-based baselines exhibit different behaviors in modeling student difficulty perception.} Conditioned Prompt LLM improves over Vanilla LLM on EEDI and SciEntsBank (UD), while Knowledge Prompt LLM shows no consistent gains and sometimes degrades performance, possibly due to long knowledge descriptions distracting the model from the core question. Under the UQ setting, both prompt-based approaches yield limited and unstable improvements, highlighting their limitations in modeling uncertainty in student simulation.


\subsection{Student Simulation Measure (RQ2)}

To evaluate whether \model can effectively simulate the behavior distribution of real students, we measure the discrepancy between responses generated by the simulated students and those from real students on the SciEntsBank dataset. Specifically, we report FID, MAUVE, and Div. KL scores, as summarized in Table~\ref{tab:student_simulation}.

\begin{table*}[htbp]
  \centering
  \caption{Performance on student simulation. Best performance is in \textbf{bold} and second best is \underline{underlined.}}
  \label{tab:student_simulation}
    \begin{tabular}{lccc|ccc}
    \toprule
    \multicolumn{1}{c}{\multirow{2}[4]{*}{Model}} & \multicolumn{3}{c}{SciEntsBank (UD)} & \multicolumn{3}{c}{SciEntsBank (UQ)} \\
\cmidrule{2-7}
          & MAUVE$\uparrow$ & FID$\downarrow$ & Div.KL$\downarrow$ & MAUVE$\uparrow$ & FID$\downarrow$ & Div.KL$\downarrow$  \\
    \midrule
    \rowcolor[rgb]{ .906,  .902,  .902} \multicolumn{7}{c}{Llama3-3B-Instruct} \\
    Vanilla LLM
      & 0.6250 & 33.54 & 12.91
      & 0.0590 & 61.59 & 16.70 \\
    Knowledge Prompt LLM
      & 0.6064 & 33.42 & 13.17
      & 0.5425 & 37.12 & 15.93 \\
    Prompt-conditioned LLM
      & 0.5475 & 35.06 & 12.93
      & 0.5580 & 35.28 & 13.48 \\
    \model(w/o target score)
      & 0.6618 & \underline{33.03} & \textbf{11.11}
      & 0.6445 & \underline{34.11} & \textbf{11.07} \\
    \rowcolor[rgb]{ .906,  .902,  .902} \multicolumn{7}{c}{Llama3-8B-Instruct} \\
    Vanilla LLM
      & 0.5392 & 35.08 & 14.22
      & 0.1878 & 59.17 & 18.31 \\
    Knowledge Prompt LLM
      & 0.6305 & 40.73 & 12.51
      & 0.5690 & 38.59 & 16.35 \\
    Prompt-conditioned LLM
      & \underline{0.6768} & 38.93 & 12.48
      & \underline{0.6730} & 37.85 & 13.76 \\
    \model(w/o target score)
      & \textbf{0.6993} & \textbf{32.91} & \underline{11.75}
      & \textbf{0.6923} & \textbf{31.64} & \underline{11.96} \\
    \bottomrule
    \end{tabular}%
  \label{tab:addlabel}%
\end{table*}%

\textbf{\model(w/o target score) generates student behavior distributions that are closer to real student data while maintaining good diversity.} In the UQ setting, it achieves high MAUVE and low FID, indicating that the method not only captures overall student behavior patterns but also reflects differences across individual students. This performance is stable across both 3B and 8B models, demonstrating the effectiveness and scalability of our approach in simulating student behavior and modeling uncertainty.

\textbf{Prompt-based methods tend to produce more homogeneous responses.} Conditioned Prompt LLM and Knowledge Prompt LLM sometimes achieve relatively high MAUVE, but often exhibit worse FID or Div.KL, suggesting that the generated student responses are concentrated in common patterns and fail to capture diverse or rare behaviors. This indicates that relying solely on prompts is insufficient to fully model student behavior distributions, particularly in capturing the full diversity of student responses.

\subsection{Student Proficiency Analysis (RQ3)}

\begin{figure}[t] 
    \centering
    \includegraphics[width=\columnwidth]{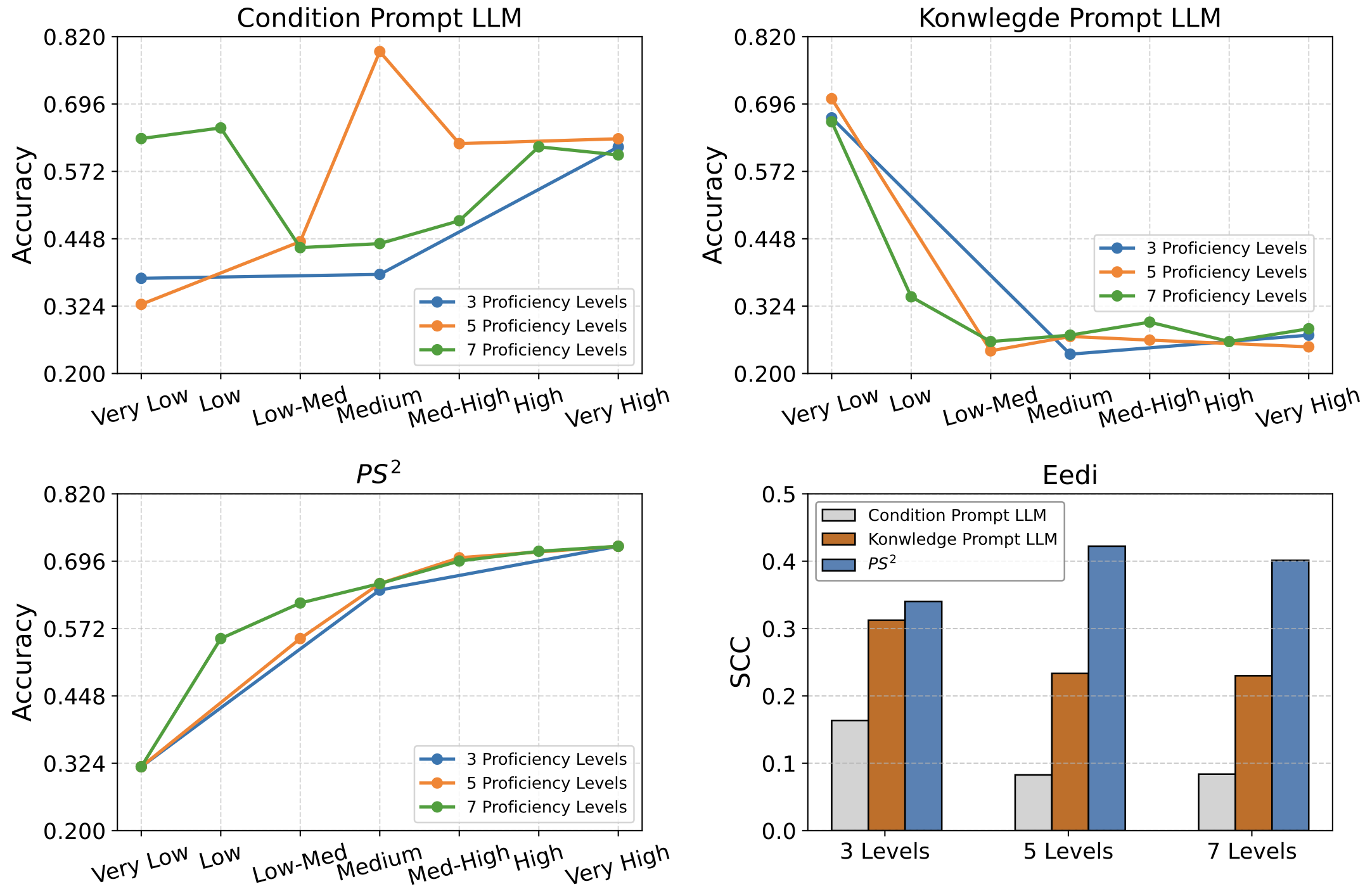}
    \caption{Accuracy of student answers under 3/5/7 proficiency levels in Eedi (top-left: Condition Prompt LLM; top-right: Knowledge Prompt LLM; bottom-left: \model). Bottom-right reports SCC for question difficulty prediction.}
    \label{fig:Student_Proficiency}
\end{figure}

To analyze the effect of increasing target proficiency levels under different granularity settings, we evaluate \model, along with {Condition Prompt LLM} and {Knowledge Prompt LLM}, on student answer accuracy and the question difficulty prediction task under 3, 5, and 7 proficiency-level settings. The results can be seen in Figure~\ref{fig:Student_Proficiency}.

Under fine-grained conditions, \model consistently preserves the correct ordering between student proficiency levels and answer accuracy. As granularity increases, \model remains capable of distinguishing students with different proficiency levels. This is attributed to \model’s interpolation between the upper- and lower-bound LLMs, enabling simulation of students with varying abilities without relying on prompts. In the question difficulty prediction task, ranking accuracy initially improves with finer granularity and only slightly declines at the highest levels, indicating stable performance. These results suggest that \model effectively captures the relationship between student proficiency and question difficulty.

\textbf{Condition Prompt LLM and Knowledge Prompt LLM perform poorly under fine-grained settings}, with lower-proficiency students sometimes achieving higher accuracy than higher-proficiency ones, indicating limited prompt control. This is especially true for Knowledge Prompt LLM: due to many knowledge points in EEDI, adding knowledge descriptions disrupts question understanding, causing the lowest-proficiency students to achieve the highest accuracy. This explains why prompt-based methods show declining ranking accuracy in question difficulty prediction as granularity increases.

\subsection{Ablation Study (RQ4)}
\begin{figure}[t] 
    \centering
    \includegraphics[width=\columnwidth]{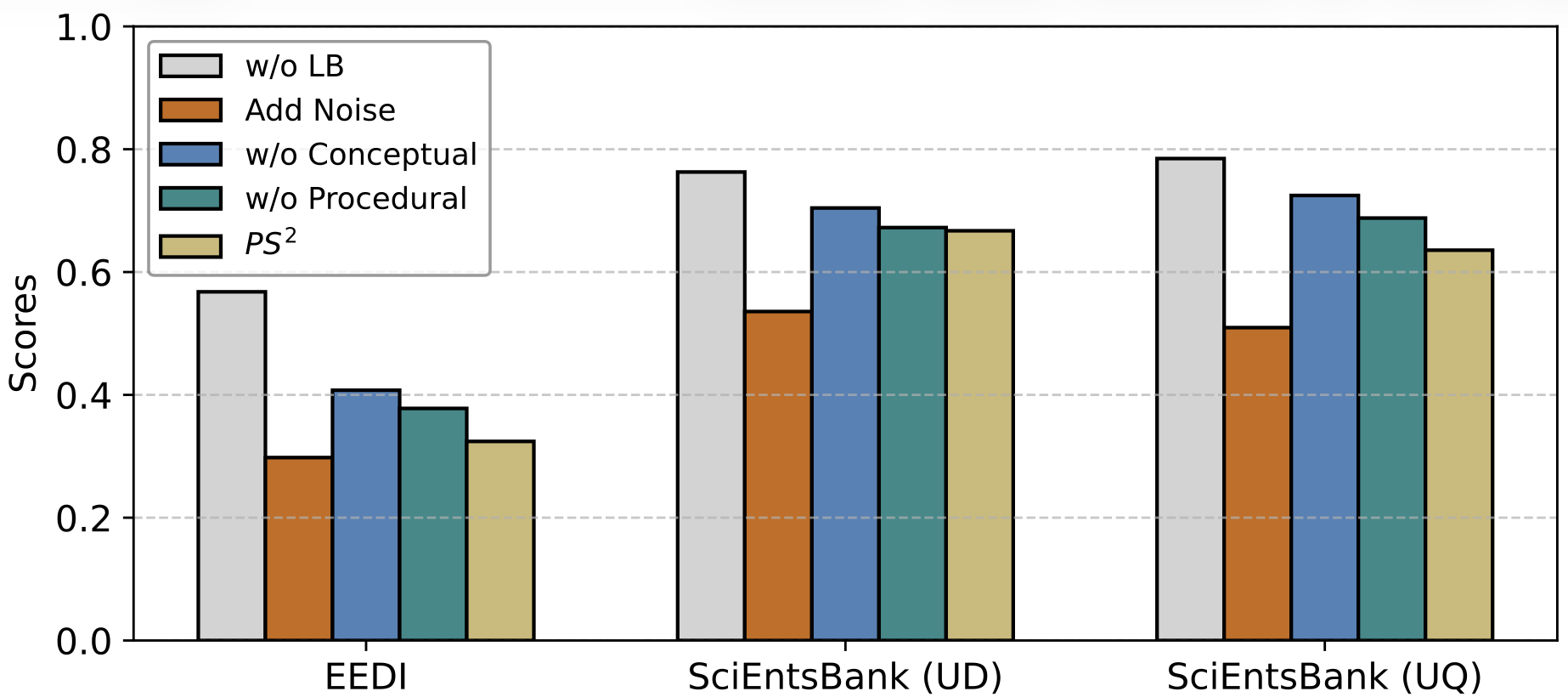}
    \caption{Scores (normalized) of different low-bound model variants across datasets.}
\label{low-bound score}
\end{figure}

To evaluate the effectiveness of lower-bound LLM training, we conduct student simulation experiments on SciEntsBank (UD, UQ) using \model and four ablated variants: (i) \textit{Add Noise}, which perturbs the upper-bound model without a lower-bound LLM; (ii)\textit{ w/o Conceptual}, which removes conceptual error; (iii) \textit{w/o Procedural}, which removes procedural error; and (iv)\textit{ w/o LB}, which disables lower-bound LLM training. We further evaluate the performance of \model and each variant’s lower-bound LLM individually. Results are reported in Table~\ref{tab:Ablation Study} and Figure~\ref{low-bound score}.

\begin{table}[htbp]
  \centering
  \caption{Ablation Study on Student Proficiency Simulation}
  \renewcommand{\arraystretch}{1.15}
  \resizebox{\linewidth}{!}{
    \begin{tabular}{lccc|ccc}
    \toprule
    \multicolumn{1}{c}{\multirow{2}[4]{*}{Model}} & \multicolumn{3}{c}{SciEntsBank (UD)} & \multicolumn{3}{c}{SciEntsBank (UQ)} \\
\cmidrule{2-7}          & \multicolumn{1}{c}{MAUVE$\uparrow$} & \multicolumn{1}{c}{FID$\downarrow$} & \multicolumn{1}{c}{Div.KL$\downarrow$} & \multicolumn{1}{c}{MAUVE$\uparrow$} & \multicolumn{1}{c}{FID$\downarrow$} & \multicolumn{1}{c}{Div.KL$\downarrow$} \\
    \midrule
    \textit{Add Noise} &   0.6792    &   44.18    &   12.07     & 0.5857 & 40.04 & \multicolumn{1}{c}{\textbf{11.27}} \\
    \textit{w/o Conceptual} & 0.6525 & \underline{33.89} & 12.01 & 0.6295 & 33.42 & 12.63 \\
    \textit{w/o Procedural} & 0.6443 & 33.96 & \underline{11.82} & 0.6023 & \underline{32.40}  & 12.17 \\
    \textit{w/o LB} & \textbf{0.7209} & 36.85 & 12.11 & \textbf{0.7113} & 33.34 & 13.74 \\
    \midrule
    \model(w/o target score)  & \underline{0.6993} & \textbf{32.91} & \textbf{11.75} & \underline{0.6923} & \textbf{31.50}  & \underline{11.96} \\
    \bottomrule
    \end{tabular}%
    }
  \label{tab:Ablation Study}%
\end{table}%

\textbf{Compared with \textit{w/o LB}, \model achieves lower FID and Div.KL while maintaining competitive MAUVE on both UD and UQ,} indicating that explicit lower-bound LLMing helps capture the diversity and deviation of real student responses. In Figure~\ref{low-bound score}, the lower-bound student LLM of \model achieves relatively low accuracy, showing that \model can reduce the performance of the lower-bound LLM without sacrificing overall performance.

\textbf{Noise injection can reduce lower-bound accuracy but sacrifices overall quality.}
Although \textit{Add Noise} achieves the lowest lower-bound accuracy in Figure~\ref{low-bound score}, its MAUVE and FID scores are significantly degraded. This indicates that simple random perturbation may encourage the lower-bound student LLM to make more errors, but it fails to learn realistic error patterns. As a result, the generated responses deviate from student distributions and exhibit lower overall quality. 


We further conduct a case study to show that, under different error generation strategies, the lower-bound student LLM produces distinct error patterns on the same question. This indicates that our error modeling strategies can effectively control the types of cognitive errors generated by the model. The detailed results are reported in Appendix~\ref{Appendix:case}.

%% file: Content/appendix.tex
\section{Model Details}
\section*{Proof of Monotonicity}
\label{Appendix:Proof}
Let $\vz^l$ and $\vz^u$ denote the logits of the lower-bound and upper-bound models at step $t$. 
For a student with proficiency level $p \in [0,1]$, we define the linearly interpolated logits as
\[
\vz(p) = (1-p)\vz^l + p \vz^u = \vz^l + p\, \Delta \vz, \quad \Delta \vz := \vz^u - \vz^l.
\]

The corresponding conditional token distributions are
\[
\pi(x \mid p) = \frac{e^{z_x(p)}}{\sum_y e^{z_y(p)}}, \quad
\pi_u(x) = \frac{e^{z_x^u}}{\sum_y e^{z_y^u}}.
\]

The KL divergence between the student model and the upper-bound model is
\[
\mathrm{KL}(\pi(p) \| \pi_u) = \sum_x \pi_x(p) \log \frac{\pi_x(p)}{\pi_u(x)}.
\]

Using the softmax derivative, we have
\begin{align*}
\frac{d \pi_x(p)}{dp}
= \pi_x(p) \Big( &\Delta z_x - \mathbb{E}_{\pi(p)}[\Delta z] \Big), \\
\mathbb{E}_{\pi(p)}[\Delta z]
&:= \sum_y \pi_y(p) \Delta z_y.
\end{align*}

Then, the derivative of the KL divergence with respect to $p$ is
\begin{align*}
\frac{d}{dp} \mathrm{KL}(\pi(p)\|\pi_u)
&= \sum_x \frac{d \pi_x(p)}{dp} \log \frac{\pi_x(p)}{\pi_u(x)} \\
&= \sum_x \pi_x(p) \Big( \Delta z_x - \mathbb{E}_{\pi(p)}[\Delta z] \Big) \log \frac{\pi_x(p)}{\pi_u(x)}.
\end{align*}

This can be written in the covariance form:
\[
\frac{d}{dp} \mathrm{KL}(\pi(p)\|\pi_u) = 
\mathrm{Cov}_{x\sim \pi(p)}\Big(\Delta z_x, \log \frac{\pi_x(p)}{\pi_u(x)}\Big) \ge 0,
\]
since $\Delta z_x$ and $\log (\pi_x(p)/\pi_u(x))$ have the same sign. Therefore, the KL divergence is monotonically increasing with respect to $p$. 

This proves that, as the proficiency level $p$ increases, the simulated student's output distribution gradually aligns with that of the upper-bound model, providing a theoretical guarantee for controlled behavior modulation.

\section{Experimental Details}
\subsection{Dataset Details}
{\textbf{EEDI} \cite{pmlr-v133-wang21a}.} EEDI is based on the NeurIPS 2020 Education Challenge dataset released by the Eedi online education platform, containing student responses to math multiple-choice questions. We extracted textual content from the original question images and excluded questions requiring visual understanding (e.g., graphs or diagrams). The resulting dataset consists of 422 unique questions and 443,433 responses from 5,533 students. We split the data into training and test sets in an 80:20 ratio. IRT was applied separately to each split to estimate question difficulty, which was then used to establish the ground-truth ranking for evaluation.

\textbf{SciEntsBank} \cite{dzikovska-etal-2013-semeval}. The SciEntsBank dataset contains nearly 11,000 short answer responses to 197 science questions across 15 domains. In our experiments, we use the 2-way labels and evaluate generalization on the Unseen Domain (UD) and Unseen Questions (UQ) splits. Since user identifiers are not provided, questions are ranked by overall accuracy as ground truth. Following ~\cite{duan2025}, we train a scoring model to assess response quality. Detailed score model training settings are provided in the Appendix~\ref{Appendix:score_model}.

\subsection{Scorer Training}

\label{Appendix:score_model}
In the SciEntsBank dataset, each student response is manually scored by trained human raters with scores in the set $\{0,1\}$. Each item includes a passage that the student must read before answering, and a reference answer is provided for scoring. To score generated student responses, we use \textbf{Qwen2.5-1.5B-Instruct} as a generative language model to construct the scoring model $R(\cdot)$. 

For each response, we combine the question text $q_i$, the item metadata $m_i$, the reference answer $a_i$, and the student response text $r_{ij}$ into a prompt $x_{ij}$, and instruct the model to output a single integer score label $y_{ij} \in \{0,1\}$. The task is formulated as a classification problem, and the model is fine-tuned using {cross-entropy loss}:

\[
\mathcal{L} = - \sum_{i,j} \sum_{c=0}^{2} \mathbf{1}[y_{ij}=c] \log \text{softmax}_c(R(c \mid x_{ij})),
\]

where $R(c \mid x_{ij})$ is the logit value output by the model for class $c$, and $\mathbf{1}[\cdot]$ is the indicator function. 

After training, the model can score responses to unseen items by {greedy decoding}, producing discrete predictions $\hat{y}_{ij}$.

\subsection{Metrics Formulation}
\label{Appendix:Formulation}
We use the following metrics to evaluate simulated student responses.

\begin{enumerate}

\item \textbf{Pearson Correlation Coefficient (PCC)} measures linear correlation between two variables $X$ and $Y$:
\begin{equation}
\text{PCC}(X,Y) = \frac{\sum_{i=1}^{n} (X_i - \bar{X}) (Y_i - \bar{Y})}{\sqrt{\sum_{i=1}^{n} (X_i - \bar{X})^2} \sqrt{\sum_{i=1}^{n} (Y_i - \bar{Y})^2}}
\end{equation}

\item \textbf{Spearman Correlation Coefficient (SCC)} measures rank correlation between $X$ and $Y$:
\begin{equation}
\text{SCC}(X,Y) = \text{PCC}(\text{rank}(X), \text{rank}(Y))
\end{equation}
Here, $\text{rank}(X)$ assigns to each element $X_i$ its position in the ascending order of $X$, with ties assigned the average rank. 

\item \textbf{Fréchet Inception Distance (FID)} between generated responses $G$ and real responses $R$ in embedding space, modeled as multivariate Gaussians $\mathcal{N}(\mu_G, \Sigma_G)$ and $\mathcal{N}(\mu_R, \Sigma_R)$:
\begin{equation}
\text{FID}(G,R) = \|\mu_G - \mu_R\|_2^2 + \mathrm{Tr}\Big(\Sigma_G + \Sigma_R - 2(\Sigma_G \Sigma_R)^{1/2}\Big)
\end{equation}

\item \textbf{MAUVE} approximates the similarity between the distributions of generated and real embeddings. Let $P$ and $Q$ be discretized distributions (clusters/buckets) of embeddings:
\begin{equation}
\text{MAUVE}(P,Q) = \max_{\alpha \in [0,1]} \text{Div}_\alpha(P,Q)
\end{equation}
where $\text{Div}_\alpha$ is the divergence curve between $P$ and $Q$.

\item \textbf{Diversity KL (Div. KL)} measures the difference in response diversity between generated and real responses. Let $S^R_{ij}$ and $S^G_{ij}$ be pairwise cosine similarities of embeddings:
\begin{equation}
\text{Div.KL} = \sum_{b} p_b^R \log \frac{p_b^R}{p_b^G}
\end{equation}
where $p_b^R$ and $p_b^G$ are the histogram probabilities of pairwise similarities in bin $b$ for real and generated responses, respectively.

\end{enumerate}

\subsection{Prompts Template}
\label{Appendix:prompt}
This subsection lists the prompt templates used throughout our pipeline.  
We use two templates to synthesize cognitive-error demonstrations for training the lower-bound student model (conceptual vs.\ procedural). We also include three question-answering templates used during evaluation, and one template for extracting item-level knowledge points.

\paragraph{Error-generation templates (for lower-bound data construction).}
We use two error-generation templates to create erroneous solutions conditioned on a specified error type.  
Each template contains (i) an error-type definition, (ii) a three-step instruction that first identifies the likely error and then produces an incorrect solution, and (iii) a style imitation block that constrains the format of the output. The placeholder \{\textit{\color{red}{StyleExample}}\} provides a reference response used only to match writing style, while \{\textit{{\color{red}Question}}\} is the target item.  
Importantly, the prompt explicitly forbids justifying mistakes, so the generated text resembles authentic student work rather than post-hoc explanations.

\begin{figure*}[t]
\centering
\begin{tcolorbox}[title={Procedural Error Prompt Template}, width=\textwidth]
Please generate an incorrect solution for a given problem based on the Error type defined below.

\textbf{Procedural Error:}

An error that occurs when a student applies knowledge, skills, or strategies incorrectly during problem-solving or task completion. Such errors may arise at different levels—routine skills, strategic operations, or higher-order control processes. Even if the student understands the underlying concepts correctly, procedural errors can lead to incorrect outcomes.

\textbf{You are supposed to:}

1. Analyze the most likely procedural error a student would make for the given problem.

2. Based on this error, generate an incorrect solution. Do not explain or justify any mistakes.

3. Provide the final student-style incorrect answer. 

\textbf{Style Imitation Instruction:}

Below is a sample response provided only to illustrate the desired writing style, structure, and level of detail.

- The sample response may be correct or incorrect.

- Do NOT assume the reasoning or answer in the sample is valid.

- Your generated response should follow a similar style and format, but must be independently generated based on the given question.

\textbf{Style Example:}

\{\textit{\color{red}{StyleExample}}\}

\textbf{Output Format:}

- Procedural Error Analysis

- Incorrect solution

- Final Answer

\textbf{Question:}

\{\textit{{\color{red}Question}}\}
\end{tcolorbox}
\end{figure*}

\begin{figure*}[t]
\centering
\begin{tcolorbox}[title={Conceptual Error Prompt Template}, width=\textwidth]
Please generate an incorrect solution for a given problem based on the Error type defined below.

\textbf{Conceptual Error: }

A student exhibits a systematic misunderstanding of relevant knowledge, concepts, principles, or relationships when solving a problem. This leads to a fundamental error in the reasoning chain. Even if the computational steps are correct, the final result is still incorrect.

\textbf{You are supposed to:}

1. Analyze the most likely procedural error a student would make for the given problem.

2. Based on this error, generate an incorrect solution. Do not explain or justify any mistakes.

3. Provide the final student-style incorrect answer. 

\textbf{Style Imitation Instruction:}

Below is a sample response provided only to illustrate the desired writing style, structure, and level of detail.

- The sample response may be correct or incorrect.

- Do NOT assume the reasoning or answer in the sample is valid.

- Your generated response should follow a similar style and format, but must be independently generated based on the given question.

\textbf{Style Example:}

\{\textit{\color{red}{StyleExample}}\}

\textbf{Output Format:}

- Procedural Error Analysis

- Incorrect solution

- Final Answer

\textbf{Question:}

\{\textit{{\color{red}Question}}\}
\end{tcolorbox}
\end{figure*}

\paragraph{Question-answering templates (for generating student responses).}
We use three templates to generate responses under different simulation settings.  
The \model and Vanilla LLM template provides only the question and serves as a baseline without explicit proficiency control.  
The Conditioned Prompt template injects a discrete student level \{\textit{{\color{red}student level}}\} to emulate common prompt-based proficiency control.  
The Knowledge Prompt template instead controls proficiency by restricting the accessible knowledge to \{\textit{{\color{red}Knowledge Description}}\}, which supports a more interpretable constraint on what the simulated student can use.

\paragraph{Knowledge-extraction template (for building the knowledge pool).}
To support the Knowledge Prompt setting, we extract up to five conceptual knowledge points from each question using a dedicated template.  
The template enforces standard academic terminology and requires JSON-only output, which reduces parsing ambiguity and enables automatic aggregation across items.

\begin{tcolorbox}[title = {\model and Vanilla LLM Question Answering Prompt Template}]
Think step by step before answering. 

\textbf{Question:}

\{\textit{{\color{red}Question}}\}
\end{tcolorbox}

\begin{tcolorbox}[title = {Conditioned Prompt LLM Question Answering Prompt Template}]
Think step by step before answering. 
You are a student of level \{\textit{{\color{red}student level}}\} (1 lowest, \{{\color{red}max level}\} highest). Answer the following question:

\textbf{Question:}

\{\textit{{\color{red}Question}}\}
\end{tcolorbox}

\begin{tcolorbox}[title = {Knowledge Prompt LLM Question Answering Prompt Template}]
Assuming you are a student, you only know the knowledge above \{\textit{{\color{red}Knowledge Description}}\}. Answer the following question:

\textbf{Question:}

\{\textit{{\color{red}Question}}\}
\end{tcolorbox}

\begin{tcolorbox}[title = {Knowledge Extraction Prompt Template}]
Extract the core knowledge points from the following question.

Rules:

- Use standard academic terms

- Focus on conceptual knowledge

- Limit to at most 5 knowledge points

- Output ONLY valid JSON, no explanation

Format:

\{
  knowledge points: [Concept 1, Concept 2]
\}

\textbf{Question:}

\{\textit{{\color{red}Question}}\}
\end{tcolorbox}

\subsection{Case Study}
\label{Appendix:case}
We present outputs from the lower-bound LLM on the same question under the \textit{w/o Conceptual} and \textit{w/o Procedural} training strategies, in order to illustrate how error generation strategies lead to distinct error patterns.

\begin{tcolorbox}[
  title=Error Pattern Example,
  colback=gray!5,
  colframe=black,
  breakable
]
\textbf{Question:} 
\(12 \div (-4) =\)

\textbf{Model Response:} 

 \textit{w/o Procedural}: ...  divides 12 by -4, {\color{red} treating -4 as a positive number}: $12\div4 = 3$.  ...

\textit{w/o Conceptual}: ... We know that {\color{red}division by a negative number is equivalent to multiplication by its absolute value}. ... So, we multiply 12 by 4. $12 \times 4 = 48$....
\end{tcolorbox}

\textbf{Analysis on \textit{w/o Procedural}.} 
The first error reflects a missing understanding of the sign rule in division: the student treats \(-4\) as \(4\) when performing the division, yielding \(3\). This indicates a systematic conceptual misunderstanding of signed arithmetic rules, where the sign of the divisor is ignored.

\textbf{Analysis on \textit{w/o Conceptual}.} 
The second error arises from an incorrect solution strategy: the student assumes that division by a negative number is equivalent to multiplication by its absolute value, and directly converts the division into multiplication, resulting in \(12 \times 4 = 48\). This constitutes a procedural error in the applied computational steps.

\subsection{Implementation Dtails}
\label{Appendix:Implementation}
\subsubsection{Knowledge Prompt LLM implementation}
\label{Appendix:Knowledge_Prompt}
The implementation of Knowledge Prompt LLM consists of three stages.

\paragraph{(1) Item-level Knowledge Extraction.}
For each training question \(q_i\), an LLM is prompted to extract up to five conceptual knowledge points in JSON format. The prompt enforces the use of standard academic terminology and restricts the output to valid JSON. The extraction is performed in parallel using a thread pool to improve efficiency. A safe JSON parser is used to handle potential code-block wrappers in the model output.

\paragraph{(2) Knowledge Aggregation.}
A raw concept-to-item mapping \(\mathcal{M}_{raw}\) is constructed, where each knowledge point is associated with all question IDs containing it:
\[
\mathcal{M}_{raw}(k) = \{i \mid k \in \mathcal{K}_i\}.
\]
The mapping is then fed back to the LLM to merge synonymous or equivalent concepts. The model returns a canonical list of knowledge points, which is deduplicated to form the final knowledge base \(\mathcal{K}_{final}\).

\paragraph{(3) Proficiency Control via Prompting.}
During inference, a target proficiency level \(p\in[0,1]\) controls the amount of knowledge provided to the simulated student. We sample a subset \(\tilde{\mathcal{K}}_p\subseteq\mathcal{K}_{final}\) such that
\[
|\tilde{\mathcal{K}}_p| = p\cdot |\mathcal{K}_{final}|.
\]
The sampled knowledge points are prepended to the student prompt. By varying \(p\), the model receives different amounts of knowledge and exhibits different proficiency behaviors.
\subsubsection{Inference Detail}
During evaluation, the vanilla LLM is decoded with a temperature of 0.7, whereas all other models employ greedy decoding.